# Application of Microgrids in Supporting Distribution Grid Flexibility

Alireza Majzoobi, *Student Member, IEEE*, and Amin Khodaei, *Senior Member, IEEE*

***Abstract*—Distributed renewable energy resources have attracted significant attention in recent years due to the falling cost of the renewable energy technology, extensive federal and state incentives, and the application in improving load-point reliability. This growing proliferation, however, is changing the traditional consumption load curves by adding considerable levels of variability and further challenging the electricity supply-demand balance. In this paper, the application of microgrids in effectively capturing the distribution network net load variability, caused primarily by the prosumers, is investigated. Microgrids provide a viable and localized solution to this challenge while removing the need for costly investments by the electric utility on reinforcing the existing electricity infrastructure. A flexibility-oriented microgrid optimal scheduling model is proposed and developed to coordinate the microgrid net load with the aggregated consumers/prosumers net load in the distribution network with a focus on ramping issues. The proposed coordination is performed to capture both inter-hour and intra-hour net load variabilities. Numerical simulations on a test distribution feeder with one microgrid and several consumers and prosumers exhibit the effectiveness of the proposed model.**

## NOMENCLATURE

*Indices:*

| | |
|---|---|
| $c$ | Superscript for distribution network consumers and prosumers. |
| ch | Superscript for energy storage charging. |
| dch | Superscript for energy storage discharging. |
| $d$ | Index for loads. |
| $i$ | Index for DERs. |
| $j$ | Index for consumers/prosumers at the distribution network. |
| $k$ | Index for intra-hour time periods. |
| $s$ | Index for scenarios. |
| $t$ | Index for inter-hour time periods. |
| $u$ | Superscript for the utility grid. |

*Sets:*

| | |
|---|---|
| D | Set of adjustable loads. |
| F | Set of flexibility constraints. |
| G | Set of dispatchable units. |
| N | Set of consumers/prosumers. |
| O | Set of operation constraints. |
| S | Set of energy storage systems. |

*Parameters:*

| | |
|---|---|
| $DR$ | Ramp down rate. |
| $DT$ | Minimum down time. |
| $E$ | Load total required energy. |
| $F(.)$ | Generation cost. |
| $MC$ | Minimum charging time. |
| $MD$ | Minimum discharging time. |
| $MU$ | Minimum operating time. |
| $UR$ | Ramp up rate. |
| $UT$ | Minimum up time. |
| $w$ | Binary islanding indicator (1 if grid-connected, 0 if islanded). |
| $\alpha, \beta$ | Specified start and end times of adjustable loads. |
| $\rho$ | Market price. |
| $\eta$ | Energy storage efficiency. |
| $\lambda$ | Value of lost load (VOLL). |
| $\psi$ | Probability of islanding scenarios. |
| $\tau$ | Time period. |
| $\Delta_1$ | Intra-hour flexibility limit. |
| $\Delta_2$ | Inter-hour flexibility limit. |
| $\Delta_1^{low}/\Delta_1^{up}$ | Microgrid time-dependent intra-hour lower/upper flexibility limit. |
| $\Delta_2^{low}/\Delta_2^{up}$ | Microgrid time-dependent inter-hour lower/upper flexibility limit. |

*Variables:*

| | |
|---|---|
| $C$ | Energy storage available (stored) energy. |
| $D$ | Load demand. |
| $I$ | Commitment state of dispatchable units. |
| $LS$ | Load curtailment. |
| $P$ | DER output power. |
| $P^M$ | Utility grid power exchange with the microgrid. |
| $SD$ | Shut down cost. |
| $SU$ | Startup cost. |
| $T^{ch}$ | Number of successive charging hours. |
| $T^{dch}$ | Number of successive discharging hours. |
| $T^{on}$ | Number of successive ON hours. |
| $T^{off}$ | Number of successive OFF hours. |
| $u$ | Energy storage discharging state (1 when discharging, 0 otherwise). |

The authors are with the Department of Electrical and Computer Engineering, University of Denver, Denver, CO 80210 USA (email: alireza.majzoobi@du.edu; amin.khodaei@du.edu).

$v$ Energy storage charging state (1 when charging, 0 otherwise).

$z$ Adjustable load state (1 when operating, 0 otherwise).

## I. Introduction

THE GROWING trend of renewable generation installations in the United States, driven primarily by current renewable portfolio standards in 27 states, efficiency incentives and net metering in 43 states, and the falling cost of renewable generation technologies [1], [2], challenges traditional practices in balancing electricity supply and demand and calls for innovative methods to reduce impacts on grid stability and reliability. Fig. 1 shows daily net load (i.e., the consumer load minus local generation) variations in California ISO, the so-called duck curve, as an example of this challenge [3]. As renewable generation increases, to reach the 33% renewable target by 2020, the power grid would require increased levels of fast ramping units to address abrupt changes (as much as 13 GW in three hours) in the net load, caused by concurrent fall in renewable generation and increase in demand.

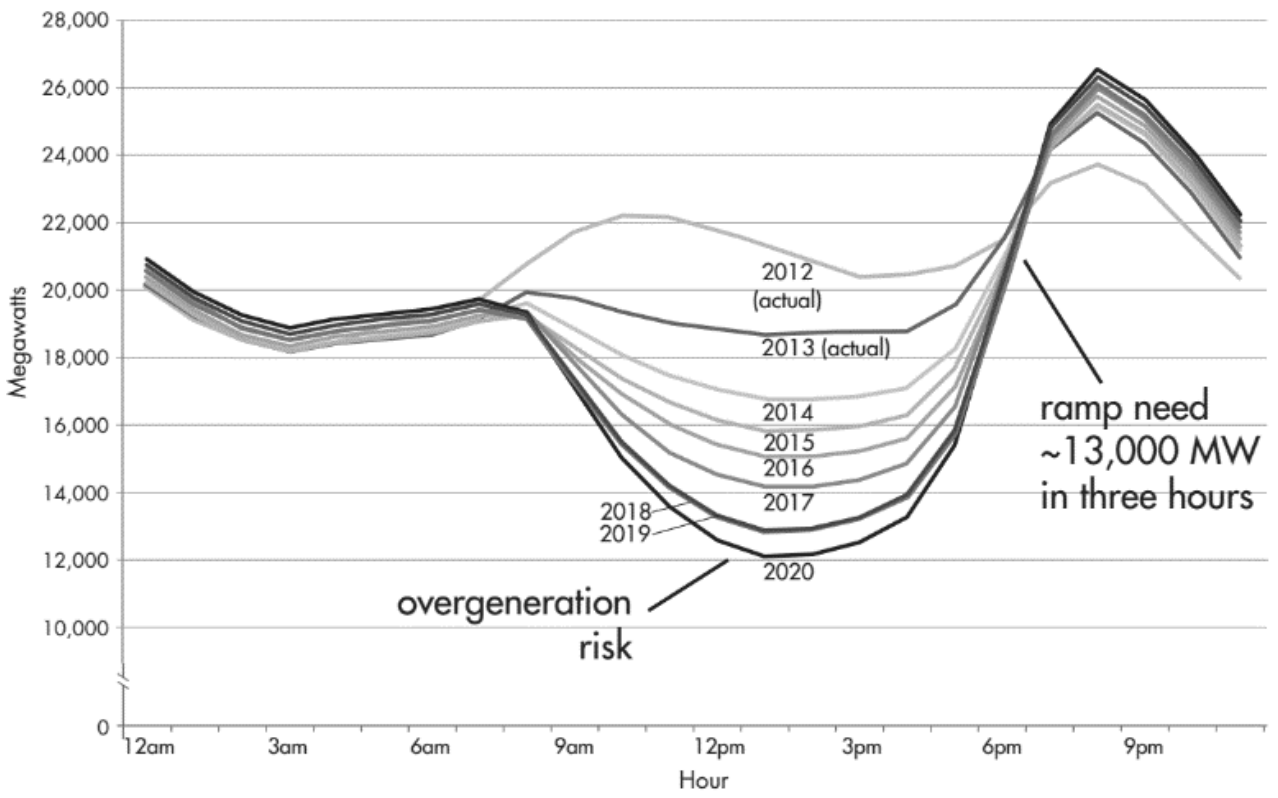

Fig. 1. The current and future estimates of over-generation and ramping effect in California [3].

To maintain system supply-demand balance, grid operators traditionally rely on bulk power generation resources, such as fast ramping hydro and thermal units, that can be quickly dispatched and ramped up. These units, however, are limited in number and capacity, capital-intensive, time-consuming to be constructed, and subject to probable transmission network congestions. Addressing the variability of renewable generation, on the other hand, has long been an attractive area of research to complement renewable generation forecasting efforts [4]. Uncertainty considerations in power system operation and planning have significantly increased in the past few years as a large amount of uncertainty sources are integrated in power systems as a result of renewable generation proliferation. The renewable generation integration problem can be investigated under two contexts of large-scale (which attempts to manage the generation of wind and solar farms) [5]-[8], and small-scale (which deals with renewable generation in the distribution level). Small-scale coordination approaches mainly focus on various methods of demand side management, such as demand response [9]-[12], energy storage [13]-[17], and aggregated electric vehicles [18], [19]. However, these methods each encounter obstacles that may prevent a viable application, such as the need for advanced metering infrastructure in deploying demand response, privacy and customer willingness issues in electric vehicle applications, and financial aspects in energy storage deployment. Leveraging available flexibility in existing microgrids for addressing renewable generation integration, as proposed in [20] and extended in this paper, will offer a potentially more viable solution to be used in distribution networks, and thus calls for additional studies. The microgrid, as defined by the U.S. Department of Energy, is “a group of interconnected loads and distributed energy resources (DER) within clearly defined electrical boundaries that acts as a single controllable entity with respect to the grid and can connect and disconnect from the grid to enable it to operate in both grid-connected or island-mode” [21]. The microgrid, as a novel distribution network architecture with local generation, control, and consumption, offers noticeable benefits to both consumers and utility companies such as enhanced reliability and resilience, reduced environmental impact, power quality improvement, improved energy efficiency by loss reduction, and network congestion relief. Microgrids can be operated in grid-connected and islanded modes. In the grid-connected mode, which is the default operation mode, the microgrid can import, export, or have zero power exchange with the utility grid to achieve the least-cost supply schedule (i.e., an economic operation). Capability to switch to the islanded mode is the salient feature of the microgrids which isolates the microgrid from faults and/or disturbances in the upstream network to achieve the least load curtailment (i.e., a reliable operation) [22]-[28]. Microgrids have been significantly deployed over the past few years and are anticipated to grow even more in the near future [29], [30], in both national and international levels [31], where future power grids can be pictured as systems of interconnected microgrids [32].

This paper focuses on the flexibility advantages of microgrids as a complementary value proposition in grid support. The microgrid capability in managing its power exchange with the utility grid in the grid-connected mode is specifically considered in this paper for mitigating the net load ramping in the distribution network and to further ensure that the power seen by the utility has manageable ramps. There have been several studies that investigate how a microgrid can participate in the upstream network market and offer services to the grid. In [33], an optimal bidding strategy via a microgrid aggregator is proposed to involve all small-scale microgrids into an electricity market via real-time balancing market bidding. In [34], an optimal bidding strategy based on two-stage stochastic linear programming for an electric vehicle aggregator who participates in the day-ahead energy and regulation markets is proposed. Furthermore, it goes on to consider market conditions and the associated uncertainty of the electric vehicle fleet. A two-stage market model for microgrid power exchange with the utility grid, via an aggregator, is proposed in [35] to achieve an efficient market equilibrium. A risk-constrained optimal hourly bidding model for microgrid aggregator is proposed in [36] to consider various uncertainties and maximize the microgrid benefit. The study in [37] proposes an optimal dispatch strategy for the residential loads via artificial neural network for calculating the demand forecast error when the demand changes are known one hour ahead with respect to the day-ahead forecasted values. The study in [38] presents a stochastic bidding strategy for microgrids participating in energy and spinning reserve markets, considering the load and renewable generation uncertainty. In [39], a stochastic look-ahead economic dispatch model for near-real-time power system operation is proposed and its benefits and implementability for assessing the power system economic risk are further explored. These works primarily rely on a market mechanism to procure microgrids' flexibility and accordingly capture the unbalanced power in the day-ahead market as well as the ramping and variabilities caused by forecast errors or unforeseen real-time events. In this paper, however, this problem is studied from a microgrid perspective, i.e., how a microgrid controller can manage local resources to offer required/desired services to the utility grid. This work is particularly important in networks that a market mechanism cannot be established but grid operators are interested in low-cost and distributed solutions in managing grid flexibility. The main contributions of this paper are listed as follows:

- A flexibility-oriented microgrid optimal scheduling model is developed to optimally manage local microgrid resources while providing flexibility services to the utility grid. This model is achieved by transforming the distribution net load variability limits into constraints on the microgrid net load.
- A coordinated grid-connected and islanded operation is considered in the model development to take into account microgrid's potential islanding while supporting the utility grid in the grid-connected mode.
- A high resolution operation is modeled via consideration of both intra-hour and inter-hour time periods, which is capable of integrating quick variations in renewable generation.

Unlike existing studies on distribution network flexibility procurement, which focus on microgrid participation in grid support via a market mechanism, this paper investigates the problem from a microgrid's perspective, i.e., how a microgrid controller can manage local resources to offer required/desired services to the utility grid.

The rest of the paper is organized as follows. Section II describes the outline of proposed flexibility-oriented microgrid optimal scheduling model. Section III develops the model formulation, including operation and flexibility constraints. Section IV presents numerical simulations to show the merits and the effectiveness of the proposed model applied to a test distribution network. Section V discusses the specific features of the proposed model, and finally, Section V concludes the paper.

## II. Model Outline

Consider a distribution feeder consisting of a set N = {1, 2, …, $N$} customers (both consumers and prosumers) and one microgrid. The net load of each customer $j \in$ N and the microgrid are respectively denoted by $P_{jtk}^{c}$ and $P_{tk}^{M}$, where $t$ is the inter-hour time index and $k$ is the intra-hour time index as demonstrated in Fig. 2.

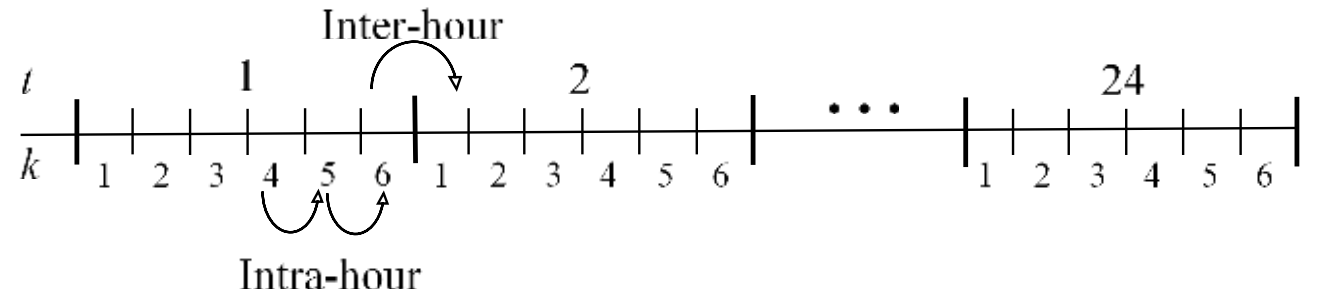

Fig. 2. The schematic diagram of inter-hour and intra-hour time intervals.

To fully supply the total net load in this feeder, a power of $P_{tk}^{u}$ needs to be provided by the utility grid where:

$$P_{tks}^{u} = P_{tks}^{M} + \sum_{j \in N} P_{jtks}^{c} \qquad \forall t, \forall k, \forall s. \tag{1}$$

To address the net load variability seen by the grid operator, the intra-hour variability (2) and inter-hour variability (3) in the utility grid power will need to be constrained:

$$\left| P_{tks}^{u} - P_{t(k-1)s}^{u} \right| \le \Delta_1 \qquad \forall t, \forall s,\ k \ne 1, \tag{2}$$

$$\left| P_{t1s}^{u} - P_{(t-1)Ks}^{u} \right| \le \Delta_2 \qquad \forall t, \forall s. \tag{3}$$

These limits are selected by the grid operator based on the day-ahead net load forecasts and desired grid flexibility during each time interval. There are various methods to determine the grid flexibility [6], [40]-[42]. If this calculated flexibility is less than the required grid flexibility, which is obtained based on net load forecasts, the grid operator can utilize distributed resources, such as microgrids, to compensate the shortage in grid flexibility. Therefore, intra- and inter-hour limits will be obtained by comparing the available and required grid flexibility. Considering the importance of grid flexibility limits on the microgrid operation, a system-level study needs to be performed by the utility company. This topic will be investigated in a follow up research. The grid operator furthermore can calculate these limits using a cost-benefit analysis, i.e., to upgrade the current infrastructure to address increasing flexibility requirements or to procure the flexibility of existing microgrids and in turn pay for their service. This topic, however, requires further analysis and modeling which will be carried out in follow up research.

Fig. 3 shows the schematic diagram of a feeder consisting of a microgrid along with other connected loads. The microgrid can be scheduled based on price considerations, i.e., local resources are scheduled in a way that the microgrid operation cost is minimized during the grid-connected mode (Fig. 3-top). The only factor impacting the microgrid scheduling results from the utility grid side is the real-time electricity price (hence the term price-based scheduling). The price-based scheduling can potentially exacerbate the consumption variability. On the other hand, microgrid resources can be scheduled in coordination with other loads in the same distribution feeder, and thus support the utility grid in mitigating potential variabilities and ensuring supply-load balance (Fig. 3-bottom). Although the objective is still to minimize the operation cost during the grid-connected mode, this scheduling is primarily based on the grid flexibility requirements (hence the term flexibility-oriented scheduling).

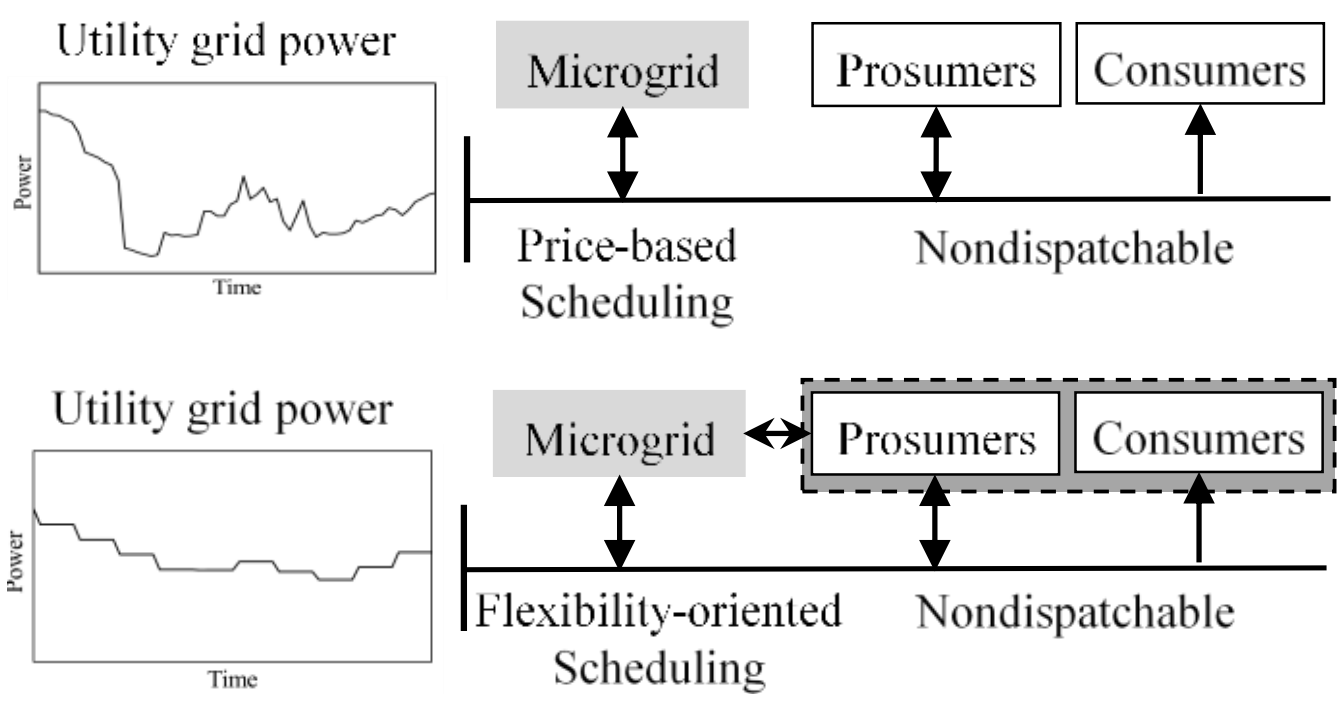

Fig. 3. Impact of the microgrid in increasing the distribution net load variabilities (top) or capturing the variabilities (bottom).

## III. Flexibility-Oriented Microgrid Scheduling – Problem Formulation

The microgrid optimal scheduling problem aims at determining the least-cost schedule of available resources (DERs and loads) while considering prevailing operational constraints, i.e.,

$$\min \sum_t \sum_k \tau \Big[\sum_{i\in G} F_i(P_{itk0}) + \rho_t^M P_{tk0}^M\Big] + \sum_t \sum_k \sum_s \tau\, \psi_s\, \lambda\, LS_{tks} \qquad (4)$$

Subject to

$$\sum_i P_{itks} + P_{tks}^M + LS_{tks} = \sum_d D_{dtks} \qquad \forall t, \forall k, \forall s, \qquad (5)$$

$$-P^{M,\max} w_{tks} \le P_{tks}^M \le P^{M,\max} w_{tks} \qquad \forall t, \forall k, \forall s, \qquad (6)$$

$$\{P_{itks}, D_{dtks}\} \in \mathrm{O}_s \qquad \forall i, \forall t, \forall k, \forall s, \qquad (7)$$

$$P_{tks}^M \in \mathrm{F}_s \qquad \forall t, \forall k, \forall s. \qquad (8)$$

The objective (4) minimizes the microgrid daily operation cost, which includes the local generation cost, cost of energy transfer with the utility grid, and the outage cost. The outage cost (also known as the cost of unserved energy) is defined as the load curtailment times the value of lost load (VOLL). The VOLL represents the customers' willingness to pay for reliable electricity service and to avoid power outages, which can also be perceived as the energy price for compensating curtailed loads. The VOLL depends on the type of customers, time and duration of outage, time of advanced notification of outage, and other specific traits of an outage. The VOLL is generally considered between \$0/MWh and \$17,976/MWh for residential customers, while for commercial and industrial customers is estimated between \$3,000/MWh and \$53,907/MWh [43, page 7]. The load balance equation (5) ensures that the sum of the injected/withdrawn power from the utility grid and local DERs (i.e., dispatchable units, nondispatchable units, and the distributed energy storage) would match the microgrid load. The load curtailment variable is used to ensure a feasible solution in the islanded operation if adequate generation is not available. The power of energy storage can be negative (charging), positive (discharging) or zero (idle). Since the power can be exchanged between the utility grid and the microgrid, $P_{tks}^M$ can be positive (power import), negative (power export) or zero. The power transfer with the utility grid is limited by (6). The binary islanding parameter (which is 1 when grid-connected and 0 when islanded) ensures that the microgrid interacts with the utility grid only during the grid-connected operation. Microgrid DERs, loads, and the main grid power transfer are further subject to operation and flexibility constraints, respectively represented by sets $\mathrm{O}_s$ and $\mathrm{F}_s$ in (7)-(8).

### A. *Operation Constraints* ($\mathrm{O}_s$)

The microgrid components to be modeled in the optimal scheduling problem include DERs (i.e., generation units and energy storage) and loads. Microgrid loads are categorized into two types of fixed (which cannot be altered and must be satisfied under normal operation conditions) and adjustable (which are responsive to price variations and/or controlling signals). Generation units in a microgrid are either dispatchable (i.e., units which can be controlled by the microgrid controller) or nondispatchable (i.e., wind and solar units which cannot be controlled by the microgrid controller since the input source is uncontrollable). The primary applications of the energy storage are to coordinate with generation units for guaranteeing the microgrid generation adequacy, energy shifting, and islanding support. From these microgrid components, only dispatchable DGs, energy storage, and adjustable loads can provide flexibility benefits for the microgrid due to their controllability. Microgrid component constraints are formulated as follows:

$$P_i^{\min} I_{it} \le P_{itks} \le P_i^{\max} I_{it} \qquad \forall i \in \mathrm{G}, \forall t, \forall k, \forall s, \qquad (9)$$

$$P_{itks} - P_{it(k-1)s} \le UR_i \qquad \forall i \in \mathrm{G}, \forall t, \forall s, k \ne 1, \qquad (10)$$

$$P_{it1s} - P_{i(t-1)Ks} \le UR_i \qquad \forall i \in \mathrm{G}, \forall t, \forall k, \forall s, \qquad (11)$$

$$P_{it(k-1)s} - P_{itks} \le DR_i \qquad \forall i \in \mathrm{G}, \forall t, \forall s, k \ne 1, \qquad (12)$$

$$P_{i(t-1)Ks} - P_{it1s} \le DR_i \qquad \forall i \in \mathrm{G}, \forall t, \forall k, \forall s, \qquad (13)$$

$$T_i^{\mathrm{on}} \ge UT_i (I_{it} - I_{i(t-1)}) \qquad \forall i \in \mathrm{G}, \forall t, \qquad (14)$$

$$T_i^{\mathrm{off}} \ge DT_i (I_{i(t-1)} - I_{it}) \qquad \forall i \in \mathrm{G}, \forall t, \qquad (15)$$

$$P_{itks} \le P_{itk}^{\mathrm{dch,max}} u_{it} - P_{itk}^{\mathrm{ch,min}} v_{it} \qquad \forall i \in \mathrm{S}, \forall t, \forall k, \forall s, \qquad (16)$$

$$P_{itks} \ge P_{itk}^{\mathrm{dch,min}} u_{it} - P_{itk}^{\mathrm{ch,max}} v_{it} \qquad \forall i \in \mathrm{S}, \forall t, \forall k, \forall s, \qquad (17)$$

$$u_{it} + v_{it} \le 1 \qquad \forall i \in \mathrm{S}, \forall t, \qquad (18)$$

$$C_{itks} = C_{it(k-1)s} - (P_{itks} u_{it} \tau / \eta_i) - P_{itks} v_{it} \tau \qquad \forall i \in \mathrm{S}, \forall t, \forall s, k \ne 1, \qquad (19)$$

$$C_{it1s} = C_{i(t-1)Ks} - (P_{it1s} u_{it} \tau / \eta_i) - P_{it1s} v_{it} \tau \qquad \forall i \in \mathrm{S}, \forall t, \forall s, \qquad (20)$$

$$C_i^{\min} \le C_{itks} \le C_i^{\max} \qquad \forall i \in \mathrm{S}, \forall t, \forall k, \forall s, \qquad (21)$$

$$T_{it}^{\mathrm{ch}} \ge MC_i (u_{it} - u_{i(t-1)}) \qquad \forall i \in \mathrm{S}, \forall t, \qquad (22)$$

$$T_{it}^{\mathrm{dch}} \ge MD_i (v_{it} - v_{i(t-1)}) \qquad \forall i \in \mathrm{S}, \forall t, \qquad (23)$$

$$D_d^{\min} z_{dtk} \le D_{dtks} \le D_d^{\max} z_{dtk} \qquad \forall d \in \mathrm{D}, \forall t, \forall k, \forall s, \qquad (24)$$

$$T_d^{on} \ge MU_d (z_{dt} - z_{d(t-1)}) \qquad \forall d \in \mathrm{D}, \forall t, \qquad (25)$$

$$\sum_{[\alpha,\beta]} D_{dtks} = E_d \qquad \forall d \in \mathrm{D}, \forall \mathrm{s}. \qquad (26)$$

Constraint (9) represents the maximum and minimum generation capacity of dispatchable units. The binary variable $I$ represents the unit commitment state which would be one when the unit is committed and zero otherwise. Dispatchable generation units are also subject to ramp up and ramp down constraints which are defined by (10)-(13). Equations (10) and (12) represent the ramping constraints for intra-hour intervals, while (11) and (13) represent the ramping constraint for inter-hour intervals. The minimum up and down time limits are imposed by (14) and (15) respectively. The minimum and maximum limits of the energy storage charging and discharging, based on the operation mode, are defined by (16) and (17), respectively. While charging, the binary charging state $v$ is one and the binary discharging state $u$ is zero; while in the discharging mode, the binary charging state $v$ is zero and the binary discharging state $u$ is one. The energy storage charging power is a negative value which is compatible with the negative amount for limitations of constraints (16) and (17) for the charging mode. Only one of the charging or discharging modes at every time period is possible, which is ensured by (18). The energy storage stored energy is calculated based on the available stored energy and the amount of charged/discharged power, which is represented in (19) and (20) for intra-hour and inter-hour intervals, respectively. The time period of charging and discharging is considered to be $\tau=(1/K)h$, where $K$ is the number of intra-hour periods and $h$ represents a time period of one hour. The amount of stored energy in energy storage is restricted with its capacity (21). The minimum charging and discharging times are represented in (22) and (23), respectively. Adjustable loads are subject to minimum and maximum rated powers (24), where binary operating state $z$ is 1 when load is consuming power and 0 otherwise. The minimum operating time (25), and the required energy to complete an operating cycle (26) are further considered for adjustable loads. It is worth mentioning that $t=0$, which would appear in (3), (14), (15), (22), and (23), represents the last hour of the previous scheduling horizon, here $t=24$.

## B. *Flexibility Constraints* ($F_s$)

Flexibility constraints represent additional limits on the microgrid power exchange with the utility grid. These constraints are defined in a way that the microgrid net load is matched with the aggregated net load of connected prosumers/consumers, so as to capture likely variations. To obtain the flexibility constraints, the value of $P^u_{tks}$, i.e., $P^u_{tks} = P^M_{tks} + \sum_{j\in N} P^c_{jtks}$, is substituted in (2) and (3). By proper rearrangements, the inter-hour and intra-hour flexibility constraints will be accordingly obtained as in (27) and (28):

$$-\Delta_1 - (\sum_j P^c_{jtk} - \sum_j P^c_{jt(k-1)}) \le P^M_{tks} - P^M_{t(k-1)s} \le \Delta_1 - (\sum_j P^c_{jtk} - \sum_j P^c_{jt(k-1)}) \qquad \forall t, \forall s, k \ne 1, \tag{27}$$

$$-\Delta_2 - (\sum_j P^c_{jt1} - \sum_j P^c_{j(t-1)K}) \le P^M_{t1s} - P^M_{(t-1)Ks} \le \Delta_2 - (\sum_j P^c_{jt1} - \sum_j P^c_{j(t-1)K}) \qquad \forall t, \forall s. \tag{28}$$

Accordingly, new time-dependent flexibility limits can be defined as follows

$$\Delta^{low}_{1,tk} = -\Delta_1 - (\sum_j P^c_{jtk} - \sum_j P^c_{jt(k-1)}) \qquad \forall t, k \ne 1, \tag{29}$$

$$\Delta^{up}_{1,tk} = \Delta_1 - (\sum_j P^c_{jtk} - \sum_j P^c_{jt(k-1)}) \qquad \forall t, k \ne 1, \tag{30}$$

$$\Delta^{low}_{2,t} = -\Delta_2 - (\sum_j P^c_{jt1} - \sum_j P^c_{j(t-1)K}) \qquad \forall t, \tag{31}$$

$$\Delta^{up}_{2,t} = \Delta_2 - (\sum_j P^c_{jt1} - \sum_j P^c_{j(t-1)K}) \qquad \forall t. \tag{32}$$

These new constraints convert the required flexibility by the grid operator to a limit on the microgrid net load. Although utility grid flexibility limits, i.e., $\Delta_1$ and $\Delta_2$, are constant and determined by the grid operator, the limits on the microgrid net load are highly variable as they comprise the aggregated net load of all $N$ customers in the distribution feeder. Depending on the considered time resolution for forecasts, these limits can change from every 1 minute to every 1 hour in the scheduling horizon. The flexibility limits can be adjusted by the grid operator to achieve the desired net load in the distribution network. For example, a value of zero for $\Delta_1$ would eliminate intra-hour variations.

It is worth mentioning that connected prosumers/consumers are considered as given parameters (forecasted) in the optimization problem. There will be no direct communications between the microgrid and the connected prosumers/consumers, where all communications will be through the grid operator. Therefore, the microgrid only communicates with the grid operator and sends/receives the required data for capturing and mitigating the distribution network net load variabilities.

## C. *Islanding Considerations*

The islanding is performed to rapidly disconnect the microgrid from a faulty distribution network, safeguard the microgrid components from upstream disturbances, and protect voltage sensitive loads when a quick solution to utility grid voltage problems is not imminent. The time and the duration of such disturbances, however, are not known to microgrids in advance. Islanding is considered in this paper via a Θ-$k$ islanding criterion, where $\Theta(=T\times K)$ represents the total number of intra-hour time periods in the scheduling horizon and $k$ represents the number of consecutive intra-hour periods that the microgrid should operate in the islanded mode. To apply this criterion to the proposed model, the binary islanding indicator $w$ is defined and added to the microgrid power exchange constraint (6). Several scenarios are defined based on the number of intra-hour time periods (for instance 144 scenarios for 10-minute intra-hour periods), and the value of $w$ in each scenario is obtained based on the Θ-$k$ islanding criterion, i.e., in each scenario $w$ will be 0 for $k$ consecutive intra-hour time periods (imposing an islanded operation) and 1 in other periods (representing the grid-connected operation). Fig. 4 shows the first five islanding scenarios, from a total of 144 scenarios, associated with a Θ-4 islanding criterion, which requires that the microgrid be able to operate in the islanded mode for any 4 consecutive intra-hour periods once it is switched to the islanded mode. Further discussions on the Θ-$k$ islanding criterion can be found in [23]. It should be noted that the proposed model is generic and can be applied to any microgrid size without loss of generality.

| *t* | 1 | | | | | | 2 | | | | | | 3 | | | | | | |
|---|---|---|---|---|---|---|---|---|---|---|---|---|---|---|---|---|---|---|---|
| *k* | 1 | 2 | 3 | 4 | 5 | 6 | 1 | 2 | 3 | 4 | 5 | 6 | 1 | 2 | 3 | 4 | 5 | 6 | ... |
| Scenario 1 | 0 | 0 | 0 | 0 | 1 | 1 | 1 | 1 | 1 | 1 | 1 | 1 | 1 | 1 | 1 | 1 | 1 | 1 | ... |
| Scenario 2 | 1 | 0 | 0 | 0 | 0 | 1 | 1 | 1 | 1 | 1 | 1 | 1 | 1 | 1 | 1 | 1 | 1 | 1 | ... |
| Scenario 3 | 1 | 1 | 0 | 0 | 0 | 0 | 1 | 1 | 1 | 1 | 1 | 1 | 1 | 1 | 1 | 1 | 1 | 1 | ... |
| Scenario 4 | 1 | 1 | 1 | 0 | 0 | 0 | 0 | 1 | 1 | 1 | 1 | 1 | 1 | 1 | 1 | 1 | 1 | 1 | ... |
| Scenario 5 | 1 | 1 | 1 | 1 | 0 | 0 | 0 | 0 | 1 | 1 | 1 | 1 | 1 | 1 | 1 | 1 | 1 | 1 | ... |

Fig. 4. First five islanding scenarios associated with a Θ-4 islanding criterion.

## IV. Numerical Simulations

A microgrid with four dispatchable units, two nondispatchable units including wind and solar, one energy storage, and five adjustable loads is used to study the performance of the proposed model. The characteristics of the microgrid DERs and loads, and the hourly market price are borrowed from [23]. The maximum ramping capability of the microgrid, based on the maximum ramping capacity of DERs, is 18 MW/h and the capacity of the line connecting the microgrid to the distribution feeder is assumed to be 10 MW. A VOLL of $10,000/MWh is considered for the microgrid.

The aggregated consumption profile of consumers/prosumers connected to the system in the same feeder as the microgrid is shown in Fig. 5. This figure consists of aggregated values for the distributed solar generation, consumption, and the net load (i.e., difference between the local consumption and generation). The net load should be supplied by the utility grid, and as the figure demonstrates, it includes considerable variabilities due to the local solar generation. The maximum ramping of this net load is 3.3 MW/10-min and the peak net load is 12.9 MW. This net load variability should be satisfied by either fast response units deployed by the utility or locally by the microgrid, where the latter is discussed here. The proposed flexibility-oriented microgrid optimal scheduling model is developed using mixed-integer programming and solved using CPLEX 12.6. It should be noted that the computation time for the studied cases was between 3 and 4 minutes, with an average of 3 min and 22 s.

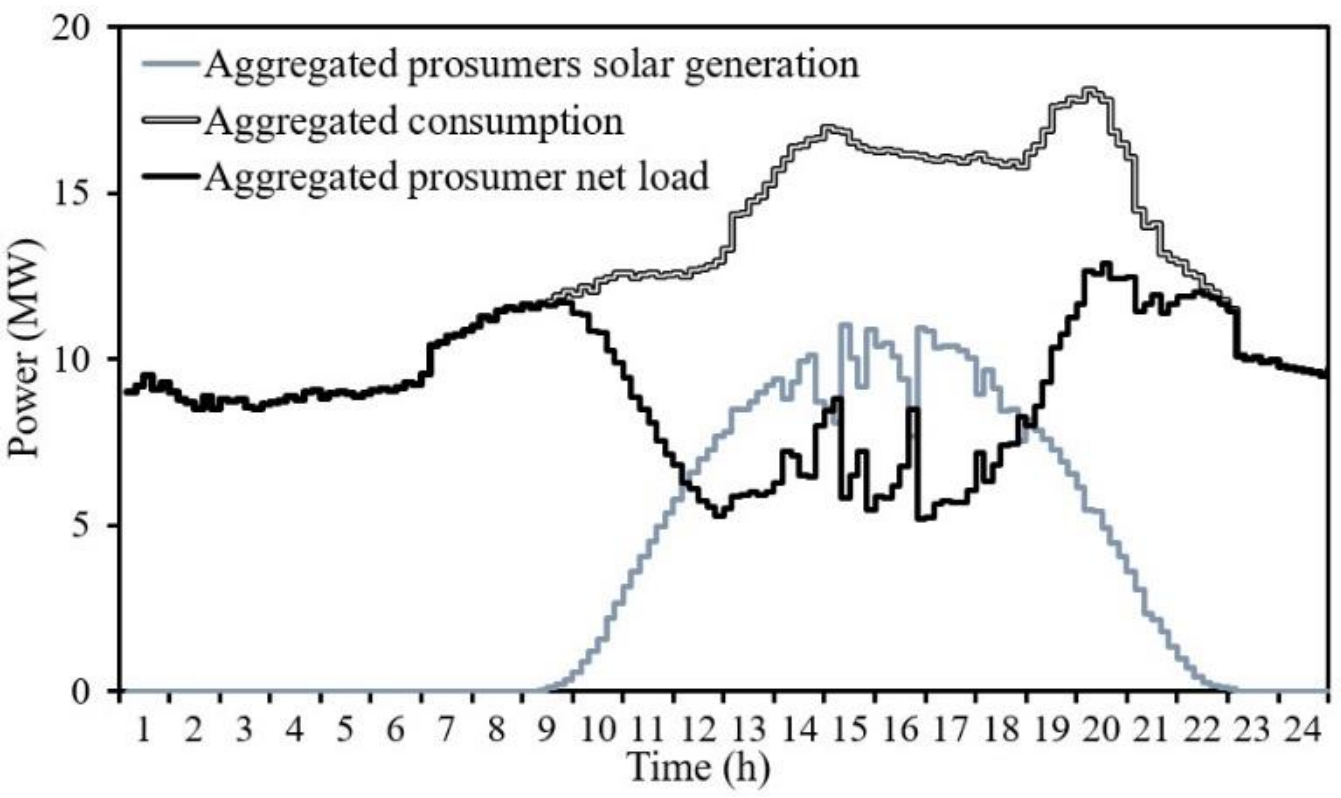

Fig. 5. Aggregated prosumers solar generation, consumption, and the net load in the distribution feeder.

**Case 1:** The grid-connected, price-based optimal scheduling is analyzed for a 24-hour horizon. The price-based scheduling denotes that the microgrid seeks to minimize its operation cost and does not have any commitment to support the utility grid in capturing distribution network net load variabilities. Table I shows the schedule of dispatchable units and the energy storage for 24 hours of operation in this case. A commitment state of 1 represents that the dispatchable unit is ON while 0 represents that the unit is not committed. The energy storage charging, discharging, and idle states are represented by -1, 1, and 0, respectively. The bold values represent changes in the schedule due to the islanding requirements. Dispatchable unit 1 has the lowest operation cost, so it is committed in all scheduling hours, while other units are committed and dispatched when required based on economic and reliability considerations. It should be noted that the amount of load curtailment during the islanded operation is considered as a measure of microgrid reliability. The energy storage is charged in low price hours and discharged in high price hours, i.e., an energy arbitrage, to maximize the benefits and minimize the operation cost. As the table shows, the islanding criterion leads to the commitment of more units in the grid-connected mode to guarantee a seamless islanding.

TABLE I
DER Schedule in Case 1

| | Hours (1-24) | | | | | | | | | | | | | | | | | | | | | | | |
|---|---|---|---|---|---|---|---|---|---|---|---|---|---|---|---|---|---|---|---|---|---|---|---|---|
| G1 | 1 | 1 | 1 | 1 | 1 | 1 | 1 | 1 | 1 | 1 | 1 | 1 | 1 | 1 | 1 | 1 | 1 | 1 | 1 | 1 | 1 | 1 | 1 | 1 |
| G2 | **1** | **1** | **1** | **1** | **1** | **1** | **1** | **1** | **1** | **1** | 1 | 1 | 1 | 1 | 1 | 1 | 1 | 1 | 1 | 1 | 1 | 1 | 1 | 1 |
| G3 | **1** | **1** | **1** | **1** | 0 | 0 | **1** | **1** | **1** | **1** | **1** | 1 | 1 | 1 | 1 | 1 | 1 | 1 | 1 | 1 | 1 | 1 | **1** | **1** |
| G4 | 0 | 0 | 0 | 0 | 0 | 0 | 0 | 0 | 0 | 0 | 0 | 1 | 1 | 1 | 0 | 1 | 1 | 1 | 1 | 1 | 1 | 1 | 0 | 0 |
| DES | -1 | -1 | -1 | -1 | -1 | **0** | **0** | 0 | 0 | 0 | 0 | 0 | 0 | 0 | 0 | 1 | 1 | 1 | 1 | 1 | 0 | 0 | 0 | 0 |

Fig. 6 depicts the microgrid net load and the distribution feeder net load (i.e., the microgrid net load plus the aggregated consumer/prosumer net load in Fig. 5). As this figure shows, the microgrid imports the power from the utility grid in low price hours and switches over to local generation when the utility grid price is high. This scheduling causes a 21.58 MW peak load for the utility grid between hours 9 and 10 (that is a new morning peak), and also exacerbates the distribution feeder ramping requirement (which is increased to 8.9 MW/10-min between hours 11 and 12 in this case). In addition, the net load variability is significantly increased in this case.

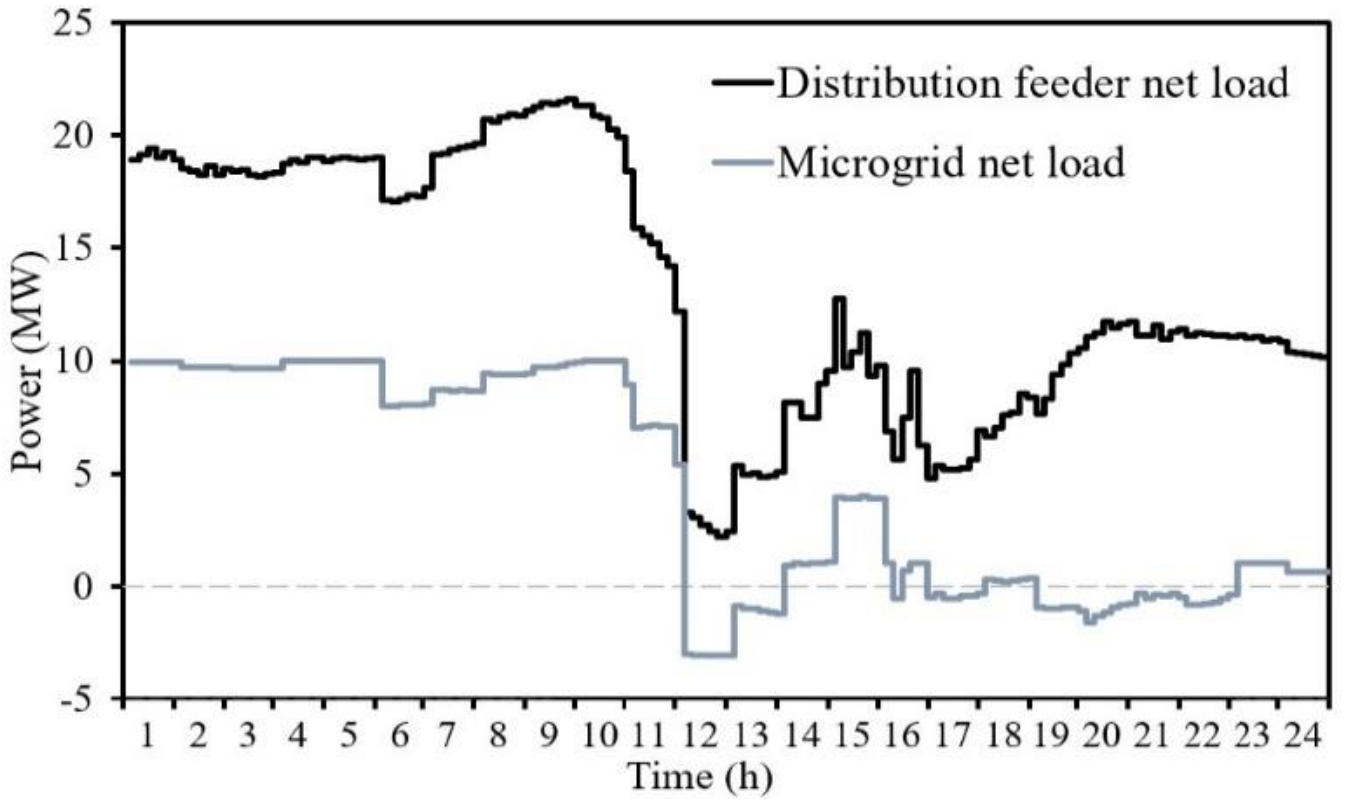

Fig. 6. Distribution feeder net load, and microgrid net load for the 24-hour horizon in Case 1.

Therefore, the utility grid encounters severe net load ramping and variations, caused by the microgrid to a great extent. This result advocates that the microgrid can potentially have a negative impact on the distribution network net load when scheduled only based on the price data and economic considerations. The microgrid operation cost in this case is $11748.3.

**Case 2:** In this case, the flexibility-oriented microgrid optimal scheduling is carried out, rather than the price-based scheduling, to support the utility grid in addressing net load variations. A $\Theta$-1 islanding criterion with 10-min intra-hour periods is considered. This islanding criterion ensures that the microgrid is capable of switching to the islanded mode to reliably supply local loads (for any 10-min islanding during the scheduling horizon), while supporting the utility grid by providing required flexibility during the grid-connected operation. The flexibility limits of 0.5 MW/10-min are considered for inter-hour and intra-hour ramping. The intra-hour and inter-hour ramping constraints are accordingly developed, as proposed in (27)-(32) and added to the developed model. Table II shows the schedule of dispatchable units and the energy storage for the scheduling horizon. The bold values represent changes in the schedule, while the highlighted cells represent changes in the dispatched power compared to Case 1. This table shows that the commitment of unit 4 and the energy storage, as well as the dispatched power of all DERs, are changed compared to Case 1 to satisfy the flexibility constraints. These changes in the schedules increase the microgrid operation cost to $12077. The difference between this cost and the microgrid operation cost in Case 1 should be paid to the microgrid, as a minimum, to incentivize the microgrid for providing flexibility and supporting the utility grid. Fig. 7 shows the distribution feeder net load and the microgrid net load in this case.

TABLE II
DER SCHEDULE IN CASE 2

| | Hours (1-24) | | | | | | | | | | | | | | | | | | | | | | | |
|---|---|---|---|---|---|---|---|---|---|---|---|---|---|---|---|---|---|---|---|---|---|---|---|---|
| G1 | 1 | 1 | 1 | 1 | 1 | 1 | 1 | 1 | 1 | 1 | 1 | 1 | 1 | 1 | 1 | 1 | 1 | 1 | 1 | 1 | 1 | 1 | 1 | 1 |
| G2 | 1 | 1 | 1 | 1 | 1 | 1 | 1 | 1 | 1 | 1 | 1 | 1 | 1 | 1 | 1 | 1 | 1 | 1 | 1 | 1 | 1 | 1 | 1 | 1 |
| G3 | 1 | 1 | 1 | 1 | 0 | 0 | 1 | 1 | 1 | 1 | 1 | 1 | 1 | 1 | 1 | 1 | 1 | 1 | 1 | 1 | 1 | 1 | 1 | 1 |
| G4 | 0 | 0 | 0 | 0 | 0 | 0 | 0 | 0 | 0 | **1** | **1** | **0** | 1 | 1 | 1 | 1 | 1 | 1 | 1 | 1 | 1 | 1 | 0 | 0 |
| DES | **0** | -1 | -1 | -1 | -1 | **-1** | **-1** | 0 | 0 | 0 | 0 | 0 | 0 | **1** | **1** | 1 | 1 | 1 | 1 | 1 | **1** | 0 | 0 | 0 |

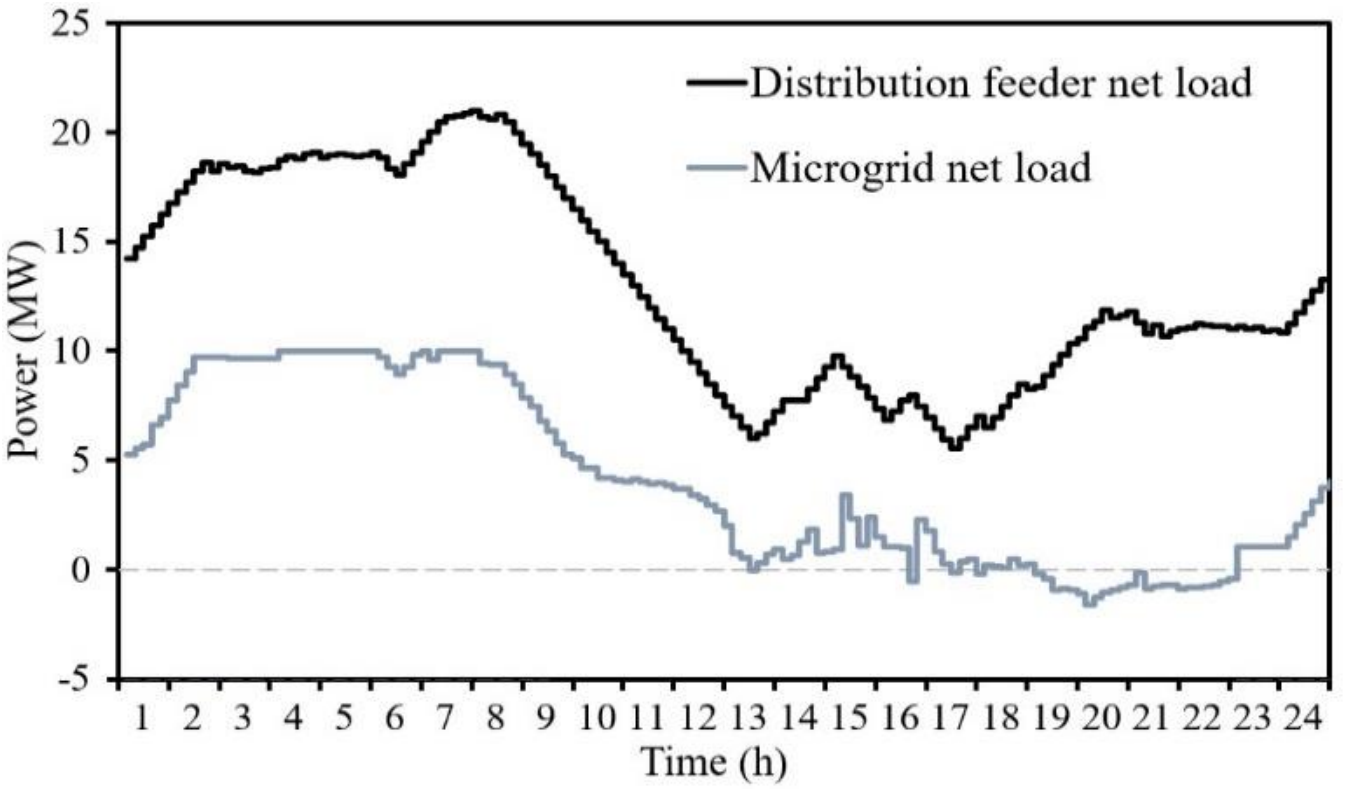

Fig. 7. Distribution feeder net load, and microgrid net load for 0.5 MW/10-min inter-hour and intra-hour utility ramping in Case 2.

Comparison of Figs. 6 and 7 shows the positive impact of the microgrid in changing the distribution network net load in a way that is desirable for the utility grid. As Fig. 6 illustrates, the distribution feeder net load, which should be supplied by the utility grid, consists of several rampings in the order of a few MW/10-min as well as a severe ramping of 8.9 MW/10-min between hours 11 and 12. In Fig. 7, however, all these variabilities are reduced to 0.5 MW/10-min as targeted by the grid operator. Moreover, Fig. 8 depicts the ramping of the utility grid in both studied cases. This figure clearly demonstrates the effectiveness of the proposed model in reducing the distribution network net load ramping, as the obtained data from Case 2 is efficiently confined between the desired ramping values.

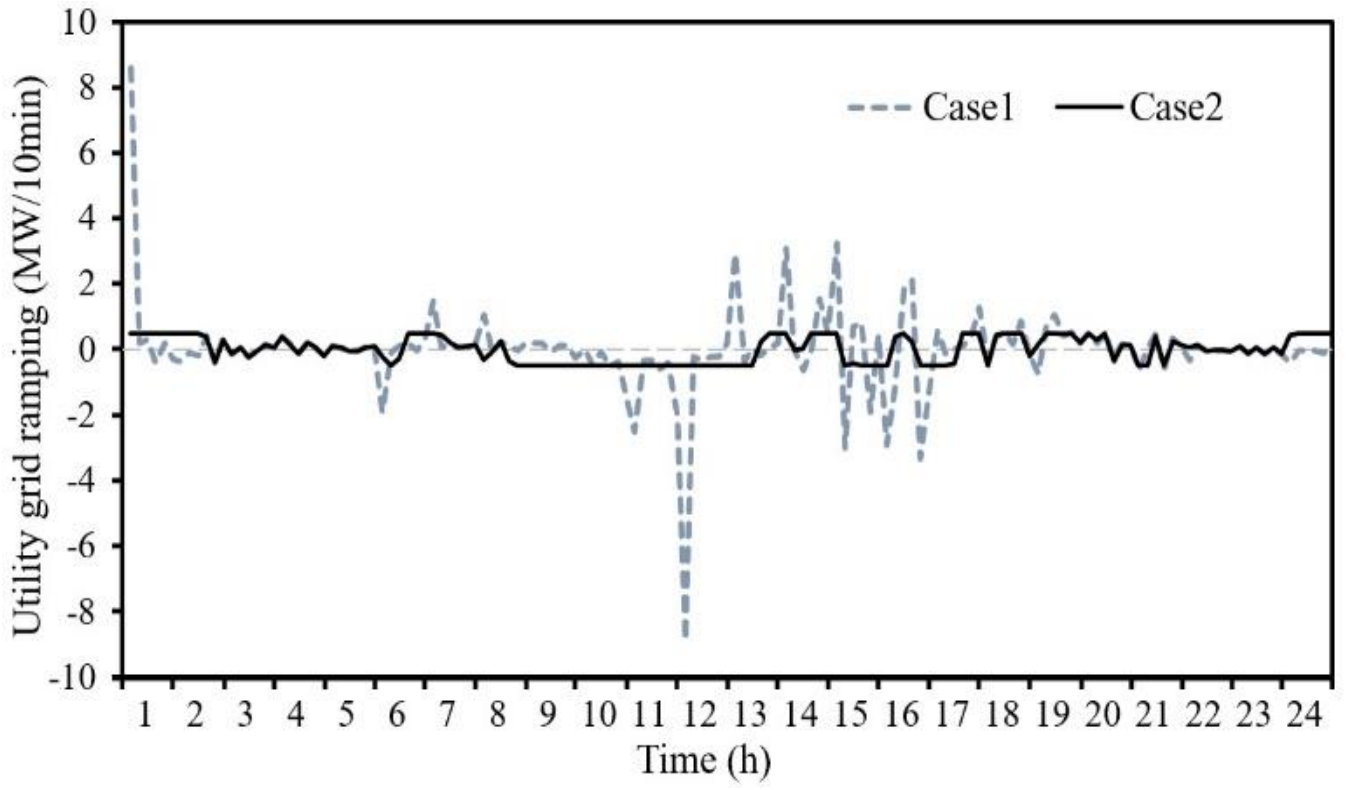

Fig. 8. Utility grid net load ramping in the two studied cases.

The results in Case 2 advocate that to obtain the desired ramping the microgrid needs to deviate from its price-based schedule. This deviation results in a $328.7 increase in the microgrid operation cost (i.e., $12077–$11748.3). This increase represents the microgrid lost revenue. To incentivize the microgrid to opt in for offering flexibility services to the utility grid, the amount of incentive that should be paid to the microgrid must be equal to or greater than this amount. If less, the microgrid would prefer to find its price-based schedule while disregarding the grid requirements. However, it would be extremely beneficial for the utility grid to incentivize the microgrid, otherwise the microgrid may exacerbate the distribution network net load variability as discussed in Case 1. It is worth mentioning that the microgrid lost revenue is a function of the consumers/prosumers net load variations as well as values of $\Delta_1$ and $\Delta_2$ which are further investigated in the following.

**Case 3:** After proving the effectiveness of the proposed model by comparing Cases 1 and 2, the impact of ramping limits is studied in this case. To show that the microgrid is also capable of meeting tight ramping limits, a value of zero is considered for the intra-hour ramping and 2 MW/10-min for the inter-hour ramping. Fig. 9 depicts the solution of this case. Considering a value of zero for intra-hour ramping completely eliminates the intra-hour variabilities in the distribution network net load, hence the obtained consumption is constant within each operation hour while it can change by up to 2 MW between any two consecutive operation hours.

To closely follow the limits, the microgrid imported power from the utility grid is decreased when the net load is increasing. Furthermore, microgrid's export to the utility grid in high price hours is changed to support the ramping limits. For instance the microgrid power export to the utility grid in hours 12-14, which was based on economic considerations, is now changed to power import from the utility grid. Fig. 10 shows the obtained results of Fig. 9 between hours 12 and 20, which better demonstrates the viable application of the microgrid in reducing the net load variability and sharp ramping.

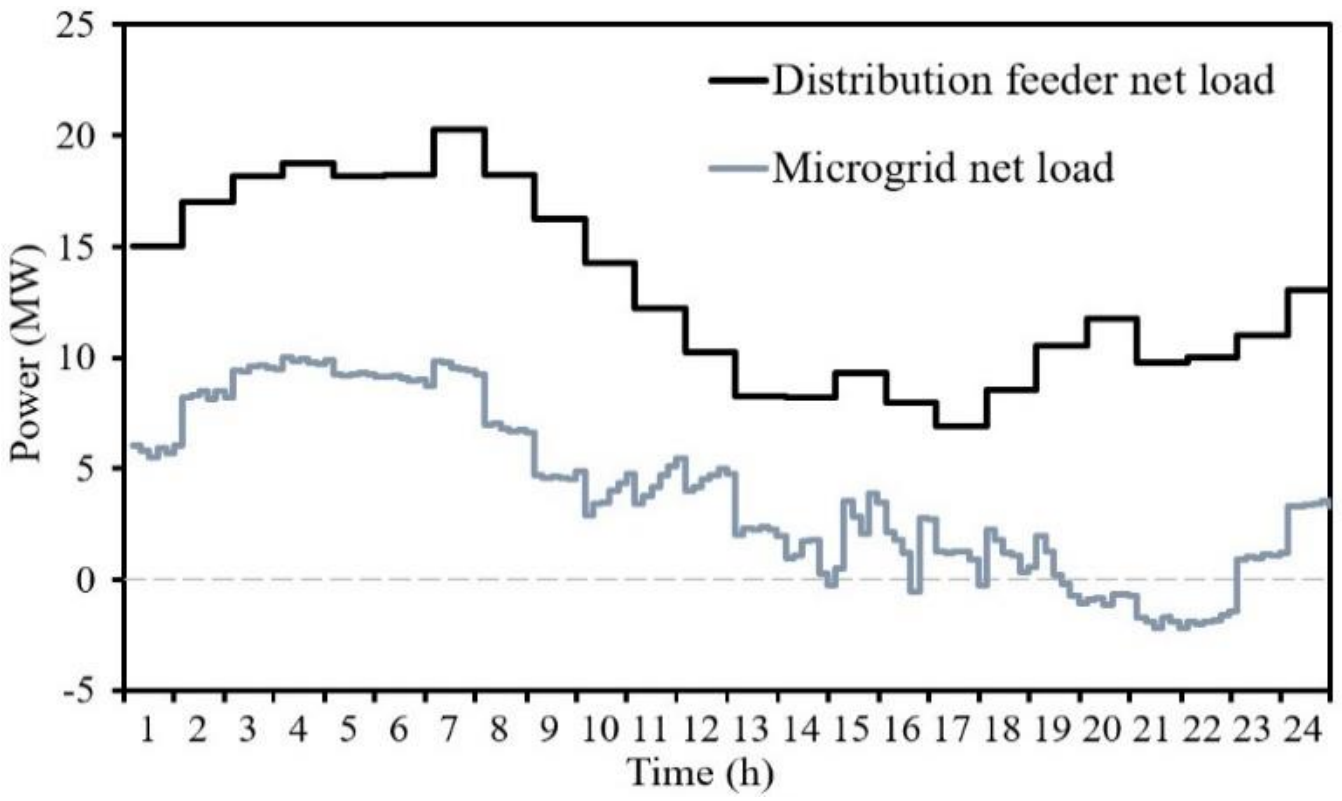

Fig. 9. Distribution feeder net load, and microgrid net load for 2 MW/10-min inter-hour and 0 MW/10-min intra-hour utility ramping.

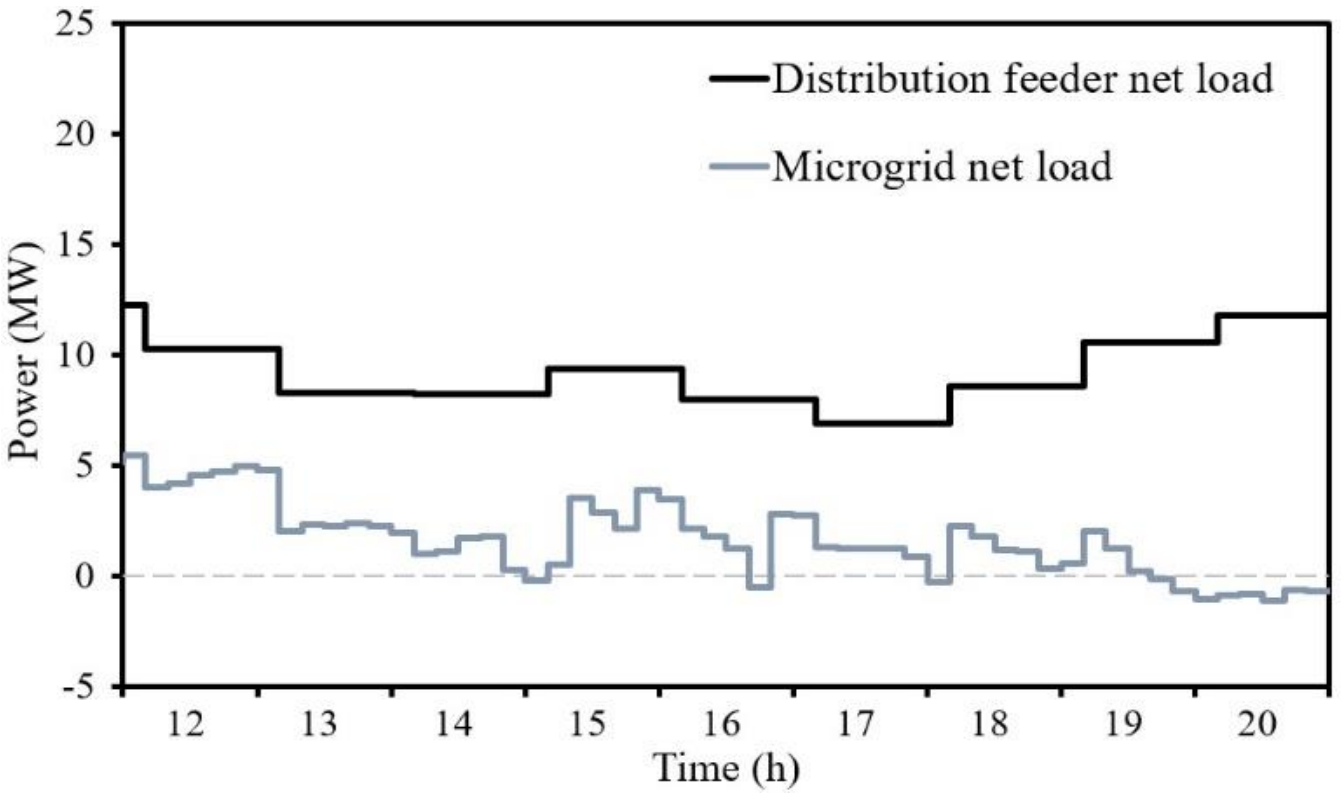

Fig. 10. Distribution feeder net load, and microgrid net load for 2 MW/10-min inter-hour and 0 MW/10-min intra-hour utility ramping, during net load peak hours.

The results of flexibility-oriented microgrid optimal scheduling for different amounts of inter-hour (changing between 0.5 and 5) and intra-hour (changing between 0 and 2) ramping limits are provided in Table III. It should be noted that all obtained results are near-optimal, mainly due to nonlinearity of the original problem and presence of uncertainties.

TABLE III
MICROGRID OPERATION COST ($) FOR VARIOUS RAMPING LIMITS

| Inter-hour ramping limit $\Delta_2$ (MW/10min) | Intra-hour ramping limit $\Delta_1$ (MW/10 min) | | | |
|---|---|---|---|---|
| | 0 | 0.5 | 1 | 2 |
| 0.5 | 36305.6 | 12077 | 11886.9 | 11825.3 |
| 1 | 19799.1 | 12011.5 | 11860.1 | 11804 |
| 1.5 | 14930.4 | 11977.5 | 11845.2 | 11796.5 |
| 2 | 13329.1 | 11951.4 | 11834.1 | 11790.1 |
| 2.5 | 12790.2 | 11936.4 | 11826.4 | 11786 |
| 3 | 12607.5 | 11925.8 | 11819.7 | 11782.1 |
| 3.5 | 12532.5 | 11916.8 | 11813.7 | 11778.6 |
| 4 | 12485.9 | 11906.8 | 11808.6 | 11775.6 |
| 4.5 | 12460.1 | 11898.3 | 11804.3 | 11772.7 |
| 5 | 12445.6 | 11891 | 11800.1 | 11770.1 |

The obtained results show that the microgrid operation cost is increased by decreasing the inter-hour and intra-hour ramping limits, however these changes are not linear. For example, the microgrid operation cost when the intra-hour ramping limit is 0 is considerably higher than other cases. This is due to two main reasons: (i) the need to commit more units and dispatch them at uneconomical operation points, in a way that they can provide the required flexibility, and (ii) the possibility of load curtailment in the microgrid. The ramping limits are added as constraints to the problem, while the load curtailment is added as a penalty to the objective function. It results in prioritizing the flexibility limit (i.e., problem feasibility) on the load curtailment (i.e., problem optimality). There of course should be additional measures to consider in order to prevent load curtailment in the microgrid which are currently under investigation by the authors. The utility grid incentive in each case must at least cover the microgrid's lost revenue. According to Table III, if the utility grid decides to eliminate the intra-hour ramping, it should pay at least $24,557.3 and $697.3 to the microgrid for $\Delta_2$ values equal to 0.5 MW/10-min and 5 MW/10-min, respectively. Whereas, in the case of 2 MW/10-min as desired intra-hour ramping, at least $77 and $21.8 should be paid to the microgrid for $\Delta_2$ equal to 0.5 MW/10-min and 5 MW/10-min, respectively. These results advocate for the importance of a cost-benefit analysis from the grid operator to determine the most suitable inter-hour and intra-hour ramping limits.

## V. DISCUSSIONS

Microgrids can potentially be utilized in distribution networks as a solution for mitigating net load ramping and variability. According to the studied cases in this paper, the following features of the proposed microgrid optimal scheduling model with multi-period islanding and flexibility constraints, could be concluded:

- Flexibility consideration: The inter-hour and intra-hour ramping constraints have been considered in the proposed model to ensure that the utility grid desired power is obtained for different time resolutions.
- Economic and reliable operation: The proposed model determines the least-cost schedule of microgrid loads and DERs while supporting the utility grid in addressing net load ramping. In addition, the consideration of $\Theta$-$k$ islanding criterion ensures the microgrid reliability in supplying local loads during the islanded mode.
- High resolution scheduling: 10-minute time interval scheduling was considered in studied cases, which offers a high resolution scheduling and is efficient for capturing net load variabilities. The proposed model offers the capability to consider various intra-hour time resolutions.
- Localized and low-cost solution: Using microgrids as local solutions for addressing distribution net load ramping can significantly reduce the utility grid investments in upgrading the generation, transmission, and distribution facilities. This significant cost saving would be made possible at the small expense of incentivizing microgrids to offer flexibility services.

## VI. CONCLUSION

In this paper, the microgrid was utilized to reduce the distribution network net load variabilities, resulted primarily due to simultaneous decrease in solar generation and increase in consumers' loads. A flexibility-oriented microgrid optimal scheduling model was proposed to efficiently schedule microgrid resources for supporting the distribution grid flexibility requirements. These flexibility requirements were considered in terms of net load ramping limits. The model was studied for intra-hour and inter-hour time intervals during the 24-hour day-ahead operation. The Θ-*k* islanding criterion was further taken into account to ensure that the microgrid has the capability to switch to the islanded mode, if needed, while supporting the utility grid during the grid-connected operation. Numerical simulations were carried out for various amounts of utility grid's desired inter-hour and intra-hour ramping to show the merits and the effectiveness of the proposed model. The results showed that the grid operator can efficiently leverage the flexibility of existing microgrids in distribution networks to address some of the most pressing flexibility-associated challenges, while removing the need for costly investments in the generation and distribution facilities.

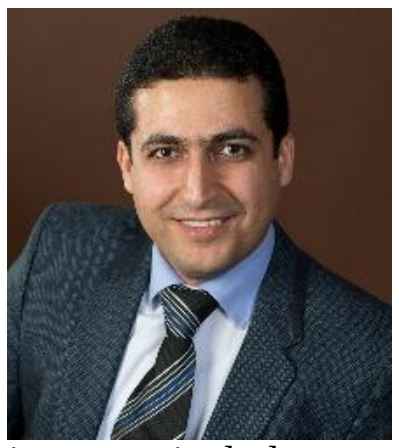

**Alireza Majzoobi** (S’15), received the B.Sc. degree from Sharif University of Technology, Tehran, Iran, in 2007, the M.Sc. degree from the University of Tehran, Tehran, Iran, in 2011, and the second M.Sc. degree from Old Dominion University, Norfolk, VA, USA, in 2015, all in electrical engineering. He is currently pursuing the Ph.D. degree in the Department of Electrical and Computer Engineering at University of Denver, Denver, CO, USA. His current research interests include power system operation and economics, microgrids, and integrating renewable and distributed resources.

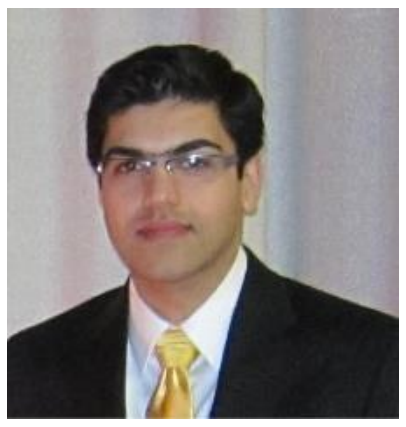

**Amin Khodaei** (SM’14) received the Ph.D. degree in electrical engineering from the Illinois Institute of Technology, Chicago, in 2010. He was a visiting faculty (2010–2012) in the Robert W. Galvin Center for Electricity Innovation at Illinois Institute of Technology. He is currently an Associate Professor in the Department of Electrical and Computer Engineering at University of Denver, Denver, CO, USA. His research interests include power system operation, planning, computational economics, microgrids, smart electricity grids, and artificial intelligence.